\documentclass[11pt]{article}
\usepackage{amsmath,amsfonts,amssymb,latexsym}

\numberwithin{equation}{section}
\def\be{\begin{equation}}
\def\ee{\end{equation}}
\def\bq{\begin{eqnarray}}
\def\eq{\end{eqnarray}}
\def\beq{\begin{eqnarray*}}
\def\eeq{\end{eqnarray*}}

\def\r{\rho}
\def\a{\alpha}
\def\b{\beta}
\def\g{\gamma}
\def\G{\Gamma}
\def\d{\delta}
\def\l{\lambda}
\def\m{\mu}

\def\t{\theta}

\begin{document}
\title{{\Huge Singularities in cosmologies with interacting fluids}}
\author{{\Large Spiros Cotsakis\footnote{\texttt{email:\thinspace
skot@aegean.gr}}\,\,\,
 and Georgia Kittou\footnote{\texttt{email:\thinspace gkittou@aegean.gr}}} \\
 School of Natural Sciences \\
University of the Aegean\\ Karlovassi 83 200, Samos, Greece}
\maketitle

\begin{abstract}
\noindent We study the dynamics near finite-time singularities of  flat isotropic universes filled with two interacting but otherwise arbitrary perfect fluids. The overall dynamical picture reveals a variety  of asymptotic solutions valid locally around the spacetime singularity. We find the attractor of all solutions with standard decay, and for `phantom' matter asymptotically at early times. We give a number of special asymptotic solutions describing universes collapsing to zero size and others ending at a big rip singularity. We also find a very complicated singularity corresponding to a logarithmic branch point that resembles a cyclic universe, and give an asymptotic local series representation of the general solution in the neighborhood of infinity.
\end{abstract}
\section{Introduction}
Fluid matter with all its ramifications has always played a key role in discussions  of cosmological singularities. In studies of the genericity of quasi-isotropic solutions \cite{l}, in studies of the structure and nature of the singularity  and energy conditions \cite{he}, in the construction of the general isotropic singularity \cite{b}, in the singularity problem of inflationary cosmology \cite{li}, or in more recent attempts towards formulating the cosmological singularity in string and brane theory \cite{le}, one sees different manifestations of the `nature abhors a vacuum' principle, i.e., using  suitable `fluids' to model the universe in its most extreme states.

In recent years there have been an increasing number of works devoted to analyzing diverse problems in situations involving  more than one cosmological fluids that show \textit{a mutual interaction} and the associated exchange of energy. Studies have been focused on a number of issues, for example a covariant description of the interaction \cite{u}, scaling solutions \cite{mi}, perturbations \cite{wa}, duality and symmetry transformations to obtain physically relevant solutions \cite{chi}, detailed solutions with energy transfer \cite{ba-cl1, ba-cl2}, and of course on  the important current issues of cosmic acceleration, dark matter and dark energy \cite{dark}.

It is therefore  important to  understand the nature of finite-time singularities that may develop in cosmological models with interacting fluids. Such an understanding will complement current studies of such models which focus on other issues and may also provide a demarcation of the range of dynamical possibilities of these models. As this issue has not, to the best of our knowledge, been pursued in a systematic way so far, it is the purpose of this paper to carry out the first steps in providing the asymptotic properties of the solutions in the neighborhood of a finite-time singularity in cosmological models with two interacting fluids. In particular, we shall focus exclusively on a flat FRW model containing two such fluids and construct asymptotic solutions which have the property to blow up at a finite-time singularity.

The asymptotic analysis of the solutions of the dynamical system  as the finite-time singularity is approached is carried out here using the method of asymptotic splittings, cf. \cite{ba-co,go}. In this method, the vector field that defines the system is asymptotically decomposed in such a way as to reveal its most important dominant features on approach to the singularity. This leads to a detailed construction of all possible local asymptotic solutions valid in the neighborhood of the finite-time singularity. These provide in turn a most accurate picture of all possible dominant features that the field possesses as it is driven to a blow up. For previous applications of this asymptotic technique to cosmological singularities, we refer to  \cite{ba-co,split}.

The plan of this paper is as follows. In the next Section, we
write the basic equations describing a flat FRW universe filled with two interacting fluids  as a dynamical system and we are lead to the asymptotic field decompositions that will yield all possible dominant features as the singularity is approached. In the following
Sections,  we present  an analysis of the asymptotic
properties of the system generally divided into power-law,
oscillatory and complete solutions.  We conclude with a discussion in the last Section, pointing into more general aspects of this problem.
\section{Asymptotic splittings}
We consider a flat FRW universe with scale factor $a(t)$ containing
two fluids with equations of state \be\label{eos} p_1=(\G
-1)\r_1,\quad p_2=(\g -1)\r_2.\ee
The unit and sign conventions we use are those of Weinberg \cite{wei}, with $8\pi G=1$. The total energy-momentum tensor is given by
\be
T^{\mathrm{\,(total)}}_{\mu\nu}=T^{\mathrm{\,(1)}}_{\mu\nu}+T^{\mathrm{\,(2)}}_{\mu\nu},
\ee
with $T^{\mathrm{\,(i)}}_{\mu\nu}=(\rho_i+p_i)u_\mu u_\nu +p_ig_{\mu\nu}, i=1,2$, with $u^\mu=\delta^\mu_0$, while we assume that the two fluids are not conserved separately, that is $T^{\mathrm{\,(1)}\,\mu\nu}_{\quad\quad;\nu}=v^\mu\neq 0$ and $T^{\mathrm{\,(2)}\,\mu\nu}_{\quad\quad;\nu}=-v^\mu$, so that $T^{\mathrm{\,(total)}\,\mu\nu}_{\quad\quad\quad;\nu}=0$. The vector $v^\mu$ describes the energy transfer between the two fluids. The time component of the conservation equation,
\be
\nabla_\nu T^{\mathrm{\,(1)}\,0\nu}=v^0,
\ee
(the spatial equations are trivial) becomes
\be
-a^3\dot p_1 +\frac{d}{dt}(a^3(\rho_1+p_1))=a^3v^0,
\ee
or,
\be
\frac{d}{da}(\rho_1 a^3)+3p_1 a^2=a^3v^0,
\ee
or, finally,
\be
\dot\rho_1+3H(\rho_1 +p_1)=Hav^0,
\ee
with $H=\dot{a}/a$ for the
Hubble expansion rate, and similarly for the second fluid.
In general, we envisage an interaction of the form
\be\label{gen}
s\equiv Hav^0=-\beta H^m\r_{1}^{\l} +\a H^n\r_2^{\,\m},
\ee
where the exponents $m,n,\l,\m$ are rational numbers indicating
that between the two fluids there is an exchange of energy that depends nonlinearly on
their densities and the Hubble rate. Depending on the signs of the
constants $\a,\b$, the fluids may `decay' to each other transferring
energy. Thus the evolution of this system is governed  by the
equations \bq\label{sys}
3H^2&=&\r_1+\r_2\nonumber\\
\dot{\r}_1+3H\G\r_{1}&=&-\beta H^m\r_{1}^{\l} +\a H^n\r_2^{\,\m}\\
\dot{\r}_2+3H\g\r_2&=&\beta H^m\r_1^\l -\a H^n\r_2^{\,\m},\nonumber
\eq
together of course with Eqs. (\ref{eos}). (The first of these equations (Friedmann equation) is obtained as in the single fluid case, cf. \cite{wei}, p. 472, but with $\rho=\rho_1+\rho_2$ and $p=p_1+p_2$ in the expression for the total energy-momentum tensor.)

Below we elaborate on the simplest
case where the exponents $m,n,\l,\m$ are all set equal to one\footnote{The previously introduced more general couplings in Eq. (\ref{gen}) correspond to perturbations of this 'standard' case. It is not \emph{a priori} obvious, however, that all such perturbations are physically relevant or distinct.}. This
case corresponds to the problem studied in \cite{ba-cl1,ba-cl2}, and it
will be interesting to compare certain of our results with theirs.
However, our  approach is completely different, the focus here being exclusively on the asymptotic approach to the
singularities of these models.
Setting all the exponents  $m,n,\l,\m=1$ and renaming $x=H$, the system (\ref{sys})
becomes equivalent to the
dynamical system
\bq
\dot x&=&y\\
\dot y&=&-Axy-Bx^3, \eq where $A=\a+\b+3\g+3\G$,
$B=3(\a\G+\b\g+3\G\g )/2.$
This defines the vector field
\be\label{vf} f(x,y)=(y,-Axy-Bx^3). \ee
The method of asymptotic
splittings developed in \cite{ba-co,go} scrutinizes all possible modes that the vector field
(\ref{vf}) attains on approach to the finite-time singularity\footnote{by a solution with a finite-time singularity we mean one where there is a time at which at least one of its components diverges. We note that the usual dynamical systems analysis through linearization etc is not relevant here, for in that one deals with equilibria, not singularities.}
located at $t=0$. These modes correspond to the different ways that
(\ref{vf}) splits as $t\rightarrow 0$. For the case we consider, the possible asymptotic modes are given by the
following three distinct \emph{decompositions}: \bq\label{decs1}
f^{(1)}&=&(y,-Axy-Bx^3),\quad\textrm{(all terms dominant case)}\\
f^{(2)}&=&(y,-Axy),\\\label{decs3} f^{(3)}&=&(y,-Bx^3). \eq
Each one
of the three decompositions (\ref{decs1})-(\ref{decs3}) into which the vector field (\ref{vf}) splits, contains
different \emph{dominant balances} that describe the precise ways into
which the dynamical system is driven asymptotically as we approach the
singularity. These balances are in general non-unique. By further
analyzing the balances of each particular decomposition, we are led
to the construction of a number of possible asymptotic formal series
valid locally in the neighborhood of the singularity, or in the
neighborhood of infinity (the latter correspond to the behaviour of the
system away from singularities, describing all possible complete
solutions). From a close examination of the form of these asymptotic
representations, we can obtain valuable information about the
genericity of the asymptotic solutions, the stability/attractor
properties of dominating solutions in the developments, and other
precise information most valuable to create a detailed shape of the
asymptotic evolution. We shall briefly comment on possible more general forms derived from the system (\ref{sys})  in the last Section.

\section{Power-law solutions, the $\delta\rightarrow 0$ attractor}
We start here our analysis of the possible asymptotic solutions towards the finite time singularity of
the system (\ref{sys}) by searching first for power-law type
solutions. The first such solution we give in this Section is the simplest and
perhaps the most important of them. Let us take the second
decomposition \be f^{(2)}=(y,-Axy), \ee and look for the possible
dominant balances, by substituting in the system $(\dot x,\dot
y)(t)=f^{(2)}$ the forms \be\label{db} x(t)=\t t^p,\quad y(t)=\xi t^q, \ee
where the coefficients $\mathbf{\Xi}\equiv (\t , \xi)\in\mathbb{C}$,
while the exponents $\mathbf{p}\equiv (p,q)\in\mathbb{Q}.$ This leads
to the unique balance \be
\mathcal{B}^{(2)}_{1}=\left[\mathbf{\Xi},\mathbf{p}\right]= \left[
(2/A,-2/A),(-1,-2)\right] ,\quad A\neq 0. \ee The candidate
subdominant part $f^{(2,\,\textrm{sub})}=(0,-Bx^3)$ of the vector
field $f^{(2)}$ satisfies \be \frac{f^{(2,\,\textrm{sub})}(\Xi\,
t^\mathbf{p})}{t^{\,\mathbf{p}-\mathbf{1}}}\equiv
\left(0,\frac{-8Bt^{3p}}{A^3t^{q-1}}\right)=\left(
0,\frac{-8\d}{A}\right). \ee Here we have utilized the
Barrow-Clifton parameter \cite{ba-cl1} \be\label{delta}
\d\equiv\frac{B}{A^2}, \ee that will play an important role in the
following. There is no way for  the vector field $f^{(2,\,\textrm{sub})}$ to be subdominant asymptotically in the sense that
\be
\frac{f^{(2,\,\textrm{sub})}(t^\mathbf{p})}{t^{\,\mathbf{p}-\mathbf{1}}}\rightarrow 0,\quad
\textrm{as}\quad t\rightarrow 0,
\ee
unless we set
\be
\d=0.
\ee
Otherwise the decomposition $f^{(2)}$ would not be acceptable asymptotically. This means that in order to satisfy this constraint, the
subdominant part has to be vanishing. We take in this case the subdominant exponent $q$ to be equal to one, cf. \cite{ba-co}.

Next we calculate the Kovalevskaya matrix ($\mathcal{K}$-matrix in
short), given by \be
\mathcal{K}=Df^{(2)}(\mathbf{\Xi})-\textrm{diag}(\mathbf{p}), \ee
where $Df^{(2)}(\mathbf{\Xi})$ is the Jacobian matrix of $f^{(2)}$,
at $\mathbf{\Xi}$, which in our case reads: \be
\mathcal{K}^{(2)}=\left(
                     \begin{array}{cc}
                       1& 1  \\
                       2 & 0 \\
                     \end{array}
                   \right).
\ee The next step is to calculate the $\mathcal{K}$-exponents for
this balance. These exponents are the eigenvalues of the
$\mathcal{K}$ matrix and constitute its spectrum,
$\textrm{spec}(\mathcal{K}^{(2)})$. The arbitrary constants of any
(particular or general) solution first appear in those terms in the
asymptotic solution series whose coefficients $\mathbf{c}_{k}$ have
indices $k=\varrho s$, where $\varrho$ is a non-negative
$\mathcal{K}$-exponent. The number of non-negative
$\mathcal{K}$-exponents equals therefore the number of arbitrary
constants that appear in the series expansions. There is always the
$-1$ exponent that corresponds to an arbitrary constant, the
position of the singularity (here at $t=0$ for notational convenience). If the balance
$\mathcal{B}^{(2)}_{1}$ is to correspond to a general solution, then it must possess  a non-negative
$\mathcal{K}$-exponent (the second arbitrary constant is the
position of the singularity). Here we find \be
\textrm{spec}(\mathcal{K}^{(2)}_{1})=\{-1,2\}, \ee so that
$\mathcal{B}_{1}$ indeed corresponds to a candidate general
solution.
Substituting the series expansions \be
x=\Sigma_{j=0}^{\infty}\,c_{j1}t^{j-1},\quad
y=\Sigma_{j=0}^{\infty}\,c_{j2}t^{j-2}, \ee in the system (\ref{sys}) and after some
manipulations to determine the coefficients of the expansions
recursively, we arrive at the following asymptotic solution around
the singularity: \be \label{expb1}
x=\frac{2}{A}t^{-1}+c_{21}t-\frac{A}{10}c_{21}^2t^3\cdots , \ee
while the $y$ expansion is obtained from the above by
differentiation. Note the arbitrary constant $c_{21}$ appearing in
this expansion signifying that this representation corresponds to a
general solution (we need two for this, the second is the arbitrary
position of the singularity).

As a final test for admission of this solution, we use the Fredholm
alternative to be satisfied by any admissible solution. This leads
to the following \emph{compatibility condition} for the positive
eigenvalue 2 and an associated eigenvector, $v_2=(1,1)$: \be
v_2^{\top}\cdot\left(\mathcal{K}-\frac{j}{s}I\right)\mathbf{c}_j=0,
\ee where $I$ denotes the identity matrix, and we have to satisfy
this at the $j=2$ level. This gives \be c_{21}=c_{22},\ee and this
is indeed true as found previously in the recursive calculation.
It follows from Eq. (\ref{expb1}) that all solutions are dominated
by the $x=H\sim \frac{2}{A}t^{-1}$ solution, that is the solution
\be\label{sol1}
H\sim \frac{2}{A}t^{-1},\quad\textrm{or}\quad a(t)\sim t^{2/A},
\ee
is an attractor of all smoothly evolving solutions at early  times, assuming the
weight-homogeneous $f^{(2)}$ decomposition asymptotically.

A comment about the results of \cite{ba-cl1} is in order. They find
that at early times the attractor solution takes the form:
\be\label{ba-cl1sol} a_{\textrm{BC}}(t)\sim
t^{-2/\sqrt{A^2-8B}},\quad\textrm{as}\quad t\rightarrow 0, \ee
whereas we find the form (\ref{sol1}). In terms of the parameter
$\d$ defined in (\ref{delta}), their solution (\ref{ba-cl1sol}) is
given by \be\label{ba-cl1sol2}
a_{\textrm{BC}}(t)\sim\left(t^{-2/|A|}\right)^{1/(1-8\d)^{1/2}},\quad\d\in
[0,1/8), \ee and we see that our solution (\ref{sol1}) includes
the $\d=0$ member of the one-parameter family of $\d$-solutions
of the form (\ref{ba-cl1sol2}). To enable the comparison, we note
that when $A<0$ in (\ref{sol1}) we have that our solution goes as $t^{-2/|A|}$,
and so the Barrow-Clifton family is asymptotic to our solution,\be
a_{\textrm{BC}}(t)\rightarrow t^{-2/|A|},\quad\textrm{as}\quad
\d\rightarrow 0, \ee (in this case, the exponent of the
$a_{\textrm{BC}}(t)$ solution   in (\ref{ba-cl1sol2}) tends to 1).
This result means that one branch of our solution (\ref{sol1}) represents a limit
function for the Barrow-Clifton family of $\delta$-solutions
(\ref{ba-cl1sol2}). Since for the validity of the $f^{(2)}$
decomposition asymptotically we were forced to take $\d =0$, we arrive at the interesting conclusion that the
Barrow-Clifton solutions (\ref{ba-cl1sol}) are all dominated by the
solution (\ref{sol1}) in this case\footnote{We will comment later on the $\d
=1/8$ limit of the $\d$-parametric family of solutions. For the
moment we note that as it is expected from (\ref{ba-cl1sol2}), as
$\d \rightarrow 1/8$ all these power-law solutions for small $t$
will  tend to zero (except of course for possible \emph{particular} exact solutions, those with a smaller number of arbitrary constants than the general solution)
and hence are expected to lose their significance
asymptotically (this is like taking the limit $\lim_{k\rightarrow
+\infty} c^{\,k}=0$, with $c\in(0,1)$).}.
\section{Phantom singularities}
Let us move on to the asymptotic analysis of the decomposition
$f^{(3)}=(y,-Bx^3)$. There are two possible balances here but only
one is  of interest for the power-law solutions of this Section (we
analyze the second balance together with other oscillatory solutions
in Section 6). Substituting in the system $(\dot x,\dot
y)(t)=f^{(3)}$ the forms (\ref{db}), we find that this balance is given by \be
\mathcal{B}^{(3)}_{1}=\left[\mathbf{\Xi},\mathbf{p}\right]= \left[
(\pm\sqrt{2/-B},\mp\sqrt{2/-B}),(-1,-2)\right] ,\quad B< 0, \ee (the
two branches give analogous results as we shall see). The candidate
subdominant part $f^{(3,\,\textrm{sub})}=(0,-Axy)$ of the vector
field $f^{(3)}$ satisfies \be \frac{f^{(3,\,\textrm{sub})}(\Xi\,
t^\mathbf{p})}{t^{\,\mathbf{p}-\mathbf{1}}}\equiv
\left(0,-A\t\xi\right)=(0,0),  \ee i.e., it vanishes only when we set $ A=0,\t
,\xi\neq 0.$ We note that the balance $\mathcal{B}^{(3)}_{1}$
corresponds to the limit \be \d\rightarrow-\infty , \ee we expect that it refers to different parts of the $\d$-family of solutions than previously\footnote{Indeed, this range of $\d$ means that the fluid parameters $\G,\g$ cannot be positive simultaneously (this is shown in detail in Ref. \cite{ba-cl1}), hence the title of this Section.}.
With the
Kovalevskaya matrix of this balance having\be
\textrm{spec}(\mathcal{K}^{(3)}_1)=(-1,4),\ee we find after some
manipulation that the series expansion corresponding to this case is
given by the form \be \label{expb3.1}
x=\pm\sqrt{\frac{-2}{B}}t^{-1}+c_{41}t^3\mp \frac{B}{12}c_{41}^2\sqrt{\frac{-2}{B}}t^7\cdots
, \ee while the $y$ expansion is obtained from the above by
differentiation.

We note here  that although the dominant term in this
expansion is the same as in (\ref{expb1}), the whole formal
expansion is a different one. The arbitrary constant $c_{41}$
appearing in the series (\ref{expb3.1}) signifies that this
representation corresponds to a general solution. Indeed, this becomes
true since the compatibility condition for the positive eigenvalue 4
(with an associated eigenvector say, $v_2=(1,3)$),  \be
v_2^{\top}\cdot\left(\mathcal{K}-\frac{j}{s}I\right)\mathbf{c}_j=0,
\ee   at the $j=4$ level gives
\be 3c_{41}=c_{42},\ee and this is  true as it follows from
the recursive calculation.

It follows from Eq. (\ref{expb3.1}) that assuming the
weight-homogeneous $f^{(3)}$ decomposition asymptotically, all
solutions dominated by the balance $\mathcal{B}^{(3)}_{1}$ (that is, those included in the family defined by (\ref{expb3.1})) are
attracted  on approach  to the singularity by the asymptotic
solution $x=H\sim \frac{2}{-B}t^{-1}$. That is, the dominating solution
\be\label{sol3.1} H\sim \frac{2}{-B}t^{-1},\quad\textrm{or}\quad
a(t)\sim t^{-2/B}, \ee is an attractor of all  smoothly evolving `phantom'
solutions at early times. Other solutions, dominated by the second balance
of this decomposition, are elucidated in the next Section.
\section{Decaying cosmologies and the borderline case}
We now focus on  the  asymptotic analysis of the all-terms-dominant
case, that is the decomposition $f^{(1)}=(y,-Axy-Bx^3)$. The
subdominant vector field is the zero field in this case, and there
are two balances:
\bq
\mathcal{B}^{(1)}_{1}&=&\left[
\left(\frac{A+\sqrt{A^2-8B}}{2B},\frac{-A-\sqrt{A^2-8B}}{2B}\right),(-1,-2)\right]\\
\mathcal{B}^{(1)}_{2}&=& \left[
\left(\frac{A-\sqrt{A^2-8B}}{2B},\frac{-A+\sqrt{A^2-8B}}{2B}\right),(-1,-2)\right].
\eq
Our analysis closely monitors the different values $\d $ may take and
we focus in this Section exclusively on power law solutions, leaving
the treatment of cyclic solutions for the next Section.
Regarding the first balance of the $f^{(1)}$ decomposition, the Kovalevskaya matrix is given by \be
\mathcal{K}^{(1)}=\left(
                     \begin{array}{cc}
                       1& 1  \\
                       -\mu+6 & -\frac{\mu}{2}+2 \\
                     \end{array}
                   \right),\quad \text{where}\quad \mu=\frac{1+\sqrt{1-8\delta}}{\delta},
\ee and we find \be
\textrm{spec}(\mathcal{K}^{(1)}_{1})=\left(-1,\frac{-\mu+8}{2}\right). \ee
As in this Section we restrict attention to power law asymptotic solutions, we examine the case $\delta=1/8$. Then we find \be
\textrm{spec}(\mathcal{K}^{(1)}_{1})=(-1,0), \ee with corresponding eigenvector \begin{equation}
\mathit{u}_2^{\text{T}}=(1,-1).
\end{equation}
The solution of the system is  particular (only one arbitrary constant, cf. \cite{ba-co,go} for this terminology) and is given by \begin{equation}\label{1/8}
x\equiv H=\frac{4}{A}t^{-1},\quad\text{or}\quad a\sim t^{4/A}.
\end{equation}
We notice the two branches of this solution, one describing universes collapsing to zero size asymptotically ($A>0$), and the other ending at a big rip singularity ($A<0$).

For the specific case $\delta=1/8$, we note the general solution found in \cite{Chim} is given by \begin{equation}\label{chimento,f1}
H^2=a^{-\frac{\text{A}}{2}}(c_3+c_4\ln a).
\end{equation}
If we set $c_4=0$, then this  is the same as the 1-parameter solution (\ref{1/8}) found above. An exact solution identical to our solution (\ref{1/8}) was also found  in \cite{ba-cl1}.

For the second balance of $f^{(1)}$ decomposition, power-law solutions can be found for $\delta=1/8$ as well as for the standard `decaying fluid' range  $0<\delta<1/8$. The Kovalevskaya matrix is given by \be
\mathcal{K}_{2}^{(1)}=\left(
                     \begin{array}{cc}
                       1& 1  \\
                       \phi+6 & \frac{\phi}{2}+2 \\
                     \end{array}
                   \right),\quad\text{where}\quad \phi =\frac{-1+\sqrt{1-8\delta}}{\delta},
\ee
with eigenvalues \be
\textrm{spec}(\mathcal{K}^{(1)}_{2})=\left(-1,\frac{\phi+8}{2}\right). \ee
Further, we notice that the  case $\delta=1/8$ of this second balance has the same eigenvalues, eigenvectors and solution as the first balance of this decomposition. These are \be \textrm{spec}(\mathcal{K}^{(1)}_{2})=(-1,0), \ee with corresponding eigenvector \begin{equation}
\mathit{u}_2^{\text{T}}=(1,-1),
\end{equation}
and solution \begin{equation}
x=\frac{4}{A}t^{-1}.
\end{equation}
Let us now turn to the behaviour of the asymptotic solutions with standard decay, that is for $0<\delta<1/8$. For definiteness,  we choose the value $\delta=1/9$. Then \be \textrm{spec}(\mathcal{K}^{(1)}_{2})=(-1,1), \ee with corresponding eigenvector \begin{equation}
\mathit{u}_2^{\text{T}}=(1,0).
\end{equation}
After further manipulations, we find that in the asymptotic expansion arbitrary coefficients are expected to be in the places $c_{11}$ and $c_{12}$, but from compatibility condition we have $c_{12}=0$, giving therefore a solution with the correct number of arbitrary constants. The final solution is a general one and reads,
\begin{equation}
x=\frac{3}{A}t^{-1}+c_{11}+\frac{A}{3}c^2_{11}t+\cdots .
\end{equation}
The dominant part of the solution asymptotically is given by \begin{equation}\label{aldb2.2}
a\sim t^{3/A},
\end{equation}  in accordance with the family found in \cite{Chim} (for $\delta=1/9$), that is \begin{equation}
[\frac{a}{a_0}]^{\frac{A}{3}}=1+C_5t+C_6t^2,
\end{equation} and it is also the member given in \cite{ba-cl1} for $\delta=1/9$.

\section{Anti-decaying, cyclic and complete universes}
This section collects together all those cases where the asymptotic solution shows a qualitatively different character than that considered so far.
The $f^{(3)}$ decomposition gives imaginary solutions for $B>0$. There are two balances:\begin{eqnarray}
\mathcal{B}^{(3)}_1&=&[(i\sqrt{2/B},-i\sqrt{2/B}),(-1,-2)]\\
\mathcal{B}^{(3)}_2&=&[(-i\sqrt{2/B},i\sqrt{2/B}),(-1,-2)].
\end{eqnarray}
Upon considering the subdominant part,  in the  case $B>0$ we find that this decomposition is asymptotically acceptable only if $A=0$, therefore when $$\delta\rightarrow\infty.$$
The eigenvalues of the Kovalevskaya matrix are for both balances given by  \be \textrm{spec}(\mathcal{K}^{(3)}_1)=\textrm{spec}(\mathcal{K}^{(3)}_2)=(-1,4), \ee with corresponding eigenvector \begin{equation}
\mathit{u}_2^{\text{T}}=(1,3).
\end{equation}
The coefficients $c_{41},c_{42}$ are expected to be arbitrary and the compatibility condition fixes one of them in terms of the other,  $3c_{41}=c_{42}$. The final solution  is given by the expansion \begin{equation}\label{f3imaginary}
x=\mp i\sqrt{2/B}\,t^{-1}+c_{41}t^3\mp i\frac{B}{12}\sqrt{2/B}c^2_{41}t^7+\cdots .
\end{equation}
This solution has $\delta\rightarrow\infty$ and so it must belong to the family of \textit{antidecaying fluids} considered in \cite{ba-cl1}. It is interesting that asymptotically the scale factor turns imaginary, perhaps an indication that the metric in this case  becomes asymptotically Euclidean. In this case we may consider as `physical' that branch of the solution that S. W. Hawking would call `compact', having zero size at the singularity, cf. \cite{haw}.

The $f^{(1)}$ decomposition, on the other hand,  leads to a very complicated singularity for $\delta>1/8$ for both remaining balances. After setting $\delta=1/2$, we get  \be \textrm{spec}(\mathcal{K}^{(1)}_1)=\textrm{spec}(\mathcal{K}^{(1)}_2)=(-1,3\mp i\sqrt{3}), \ee with corresponding eigenvectors \begin{equation}
\mathit{u}_2^{\text{T}}=(1,2\mp i\sqrt{3}).
\end{equation}
For both balances, the second eigenvalue of the $\mathcal{K}$-matrix has positive real part. The solution of the system then reads \begin{equation}\label{aldim}
x=\frac{1\pm i\,\sqrt{3}}{A}t^{-1},
\end{equation}
or, in terms of the scale factor we find\begin{equation}\label{aldimsc.f}
a\sim t^{\frac{1\pm i\sqrt{3}}{A}}.
\end{equation}
This defines a multifunction on approach to the $t=0$ singularity which is obviously a logarithmic branch point admitting no Puiseux series representation. In this case, the scale factor never returns to its original value no matter how many times $t$ loops around zero.

Lastly, we give another sort of solution.
When all the eigenvalues of the Kovalevskaya matrix are negative, the solution escapes away from the singularity towards infinity. For the $f^{(1)}$ decomposition of the system, such a state appears when we examine the first balance with $0<\delta<1/8$. If we choose $\delta=1/9$, then   \be \textrm{spec}(\mathcal{K}^{(1)}_1)=(-1,-2), \ee with a corresponding eigenvector \begin{equation}
\mathit{u}_2^{\text{T}}=(1,-3).
\end{equation}
To construct a suitable expansion for this case, we need to take the multiplicative inverse\footnote{that is the least common multiple of the set of subdominant exponents and the positive Kovalevskaya exponents, cf. \cite{ba-co,go}.} to be equal to one,  $s=-1$, and the coefficients $c_{21},c_{22}$ are arbitrary. Following the method of asymptotic splittings, we are led to the compatibility condition $3c_{21}=-c_{22}$,  and finally the asymptotic solution valid in the neighborhood of infinity: \begin{equation}\label{f1negative}
x=\frac{6}{A}t^{-1}+c_{11}t^{-2}+c_{21}t^{-3}+\cdots .
\end{equation} This is a general solution valid away from any finite time singularity, showing a standard decay between the two fluids.
\section{Discussion}
In this paper we provided a demarcation of the singular phenomena that emerge when we consider two interacting perfect fluids in a flat FRW universe. We have examined what happens when we take this system asymptotically to a finite-time singularity. We have found a number of regimes described by different asymptotic solutions - seven different behaviours in all.

There is an asymptotic solution that acts as an attractor, a limit function to a wide family of solutions parametrized by the parameter $\d$. This solution is a member of a family of singular asymptotes that has the same number of arbitrary functions as the general solution, and attracts all these smoothly evolving solutions at early times in the `direction' $\d\rightarrow 0$. There is an analogous behaviour for the so-called `phantom' regime of asymptotic solutions. There are also decaying solutions collapsing to zero size, and decaying solutions to a big rip singularity, but these are of less generality than the afore-mentioned behaviour, valid for special values of $\d$. The general solution towards the singularity with `standard decay' (that is in the range $\d\in (0,1/8)$) was also picked by our asymptotic method, and it was constructed for a concrete  parameter value. We found solutions of the `antidecaying' type that approach the finite-time singularity turning purely imaginary in the parameter limit $\d\rightarrow\infty$, these are perhaps more amenable to a quantum cosmological description. We also gave a very peculiar solution having a log-type branch point singularity describing a `cyclic' universe at 'early' times. Lastly, we have given the behaviour of solutions away from singularities and towards infinity.

The existence of the singular  behaviours unraveled in this paper makes the dynamics of  cosmologies with two interacting fluids especially interesting on approach to their singularities, and the singularity in such models deserves to be further studied. One aspect of the problem that is currently under study is whether these forms of approach to the interacting fluid singularities are stable to perturbations of the $m,n,\l,\m$ exponents away from the value one we considered in this work. This may demand a reformulation of the problem using more suitable variables. Another important issue that is also under examination is precisely how the inclusion of curvature alters the behaviours found in the flat case and whether new and distinct forms are possible. We plan to return to these more involved issues in the future.
\section*{Acknowledgements}
We thank Manos Saridakis, Luis Urena and David Wands  for discussions and useful comments, and an anonymous referee for useful suggestions. The work of G.K. is supported by a PhD grant co-funded by the European Union (European Social Fund-ESF) and national resources under the framework `Herakleitus II: Action for the enforcement of human research potential through the realization of doctorate work' which is gratefully acknowledged.

\end{document}